\documentclass{svjour3}

\smartqed  
\usepackage{graphicx}
\usepackage{amsmath}

\newcommand{\R}{\mathcal{R}}

\begin{document}

\title{The Structure of Autocatalytic Sets:\\
       Evolvability, Enablement, and Emergence}
\titlerunning{The Structure of Autocatalytic Sets}
\author{Wim Hordijk \and Mike Steel \and Stuart Kauffman}
\institute{
  Wim Hordijk \at
  SmartAnalytiX.com \\
  \email{wim@SmartAnalytiX.com}
  \and
  Mike Steel \at
  Biomathematics Research Centre \\
  University of Canterbury \\
  Christchurch, New Zealand \\
  \email{m.steel@math.canterbury.ac.nz}
  \and
  Stuart Kauffman \at
  University of Vermont \\
  Burlington, VT, USA \\
  and \\
  Tampere University of Technology \\
  Tampere, Finland \\
  \email{stukauffman@gmail.com}
}
\date{Received: date / Accepted: date}
\maketitle

\begin{abstract}
This paper presents new results from a detailed study of the structure of autocatalytic sets. We show how autocatalytic sets can be decomposed into smaller autocatalytic subsets, and how these subsets can be identified and classified. We then argue how this has important consequences for the evolvability, enablement, and emergence of autocatalytic sets. We end with some speculation on how all this might lead to a generalized theory of autocatalytic sets, which could possibly be applied to entire ecologies or even economies.
\end{abstract}

\section{Introduction}

Origin of life research seems divided between two paradigms: {\it genetics-first} and {\it metabolism-first}. However, a common theme in both is the idea of {\it autocatalysis}. In fact, a third alternative, that of {\it collectively autocatalytic sets}, was introduced more or less independently several times \cite{Kauffman:71,Eigen:77,Dyson:82}, and used in later origin of life models \cite{Wachterhauser:90,Ganti:97,Rosen:91,Letelier:06}. Furthermore, recent experimental advances in creating such sets in a laboratory setting \cite{Sievers:94,Ashkenasy:04,Hayden:08,Taran:10} have sparked a renewed interest in autocatalytic sets.

In this paper, we continue our own (theoretical) studies of autocatalytic sets. In previous work \cite{Kauffman:71,Kauffman:86,Kauffman:93,Steel:00,Hordijk:04,Mossel:05,Hordijk:10,Hordijk:11,Hordijk:12}, we used a mathematical framework to investigate the probability of existence of autocatalytic sets under various conditions (model parameters), to answer questions about the required level of catalysis needed for autocatalytic sets to emerge, and we considered various model variants. We also compared and contrasted our own work with other, related models and methods (which we will not repeat here, but see, e.g., \cite{Hordijk:10}), and presented results that complement, but also go beyond what has been reported elsewhere so far.

Here, we present new results from a detailed study of the actual {\it structure} (as opposed to merely the existence) of autocatalytic sets. In particular, we show empirically, theoretically, and through an illustrative example, that autocatalytic sets can often be decomposed into smaller subsets which themselves are autocatalytic, and how these subsets can be identified and classified. We then argue that this structural decomposition of autocatalytic sets has important consequences for their potential {\it evolvability}, how they can {\it enable} their own growth and also the coming into existence of other autocatalytic (sub)sets, and how this can possibly give rise to higher-level, {\it emergent} structures. Finally, we end the paper with some provoking but plausible ideas of how the theory of autocatalytic sets, having started in the context of the origin of life, might be generalized to a theory of functional organization and emergence. Such a generalized theory of autocatalytic sets could, as we hope and envision, perhaps even be applicable to ecology and economics.

\section{Chemical reaction systems and autocatalytic sets}

In previous work we introduced and studied a formal model of chemical reaction systems and autocatalytic sets in the context of the origin of life problem \cite{Kauffman:71,Kauffman:86,Kauffman:93,Steel:00,Hordijk:04,Mossel:05,Hordijk:10,Hordijk:11,Hordijk:12}. Here, we will only briefly review the relevant definitions and results.

A {\it chemical reaction system} (CRS) is defined as a tuple $Q=\{X,\R,C\}$ consisting of a set of molecule types $X$, a set of reactions $\R$ (where each reaction transforms a set of reactants into a set of products), and a catalysis set $C$ indicating which molecule types catalyze which reactions. We also consider the notion of a food set $F \subset X$, which is a subset of molecule types that are assumed to be freely available from the environment. In one particular model of a CRS, known as the binary polymer model \cite{Kauffman:71,Kauffman:86,Kauffman:93}, molecule types are represented as bit strings up to a certain length $n$, reactions are simply ligation (``gluing'' two bit strings together into one longer one) and cleavage (splitting one bit string into two shorter ones), and catalysis is assigned at random according to some parameter $p$ (the probability that a molecule type catalyzes a reaction). The food set consists of all molecule types up to a certain length $t \ll n$.

\begin{figure}[htb]
\centering
\includegraphics[scale=.7]{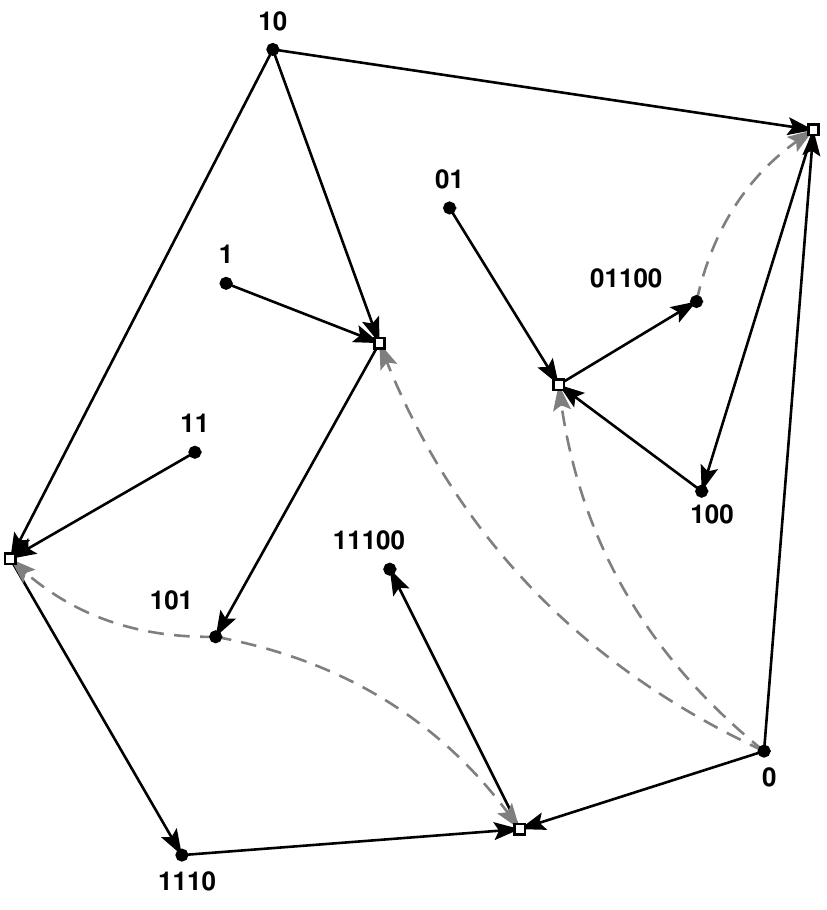}
\caption{An example of an RAF set that was found by the RAF algorithm in an instance of the binary polymer model. Molecule types are represented by black dots and reactions by white boxes. Solid arrows indicate reactants and products coming in and out of a reaction, while dashed arrows indicate catalysis. The food set is $F=\{0,1,00,01,10,11\}$.}
\label{fig:RAF_1}
\end{figure}

Informally, an {\it autocatalytic set} (or RAF set, in our terminology) is now defined as a subset $\R' \subseteq \R$ of reactions (and associated molecules types) in which:
\begin{enumerate}
\item each reaction $r \in \R'$ is catalyzed by at least one molecule type involved in $\R'$;
\item all reactants and catalysts in $\R'$ can be created from the food set $F$ by using reactions only from $\R'$ itself.
\end{enumerate}
A formal definition is provided in \cite{Hordijk:04,Hordijk:11}, where we also introduced a polynomial-time (in the size of the reaction set $\R$) algorithm for finding RAF sets in a general CRS $Q=\{X,\R,C\}$. Figure \ref{fig:RAF_1} shows an example of an RAF set that was found by our algorithm in an instance of the binary polymer model with parameter values $n=5$, $t=2$, and $p=0.0045$.

In \cite{Hordijk:04} we showed computationally, and then proved theoretically in \cite{Mossel:05}, that only a linear growth rate in the level of catalysis $f$ (i.e., the average number of reactions catalyzed per molecule type) with the size of the largest molecules $n$, suffices to get RAF sets with high probability in instances of the binary polymer model. Furthermore, the level of catalysis only needs to be roughly $1 < f < 2$, i.e., between one and two reactions catalyzed per molecule (for $n$ at least up to 20), which is a chemically realistic number. In \cite{Hordijk:11,Hordijk:12} we studied a variant of the binary polymer model where catalysis is based on a more realistic template-matching constraint. However, the main results from the basic model remain the same and, moreover, can be mathematically predicted \cite{Hordijk:12}.

Finally, we note that the RAF sets that are found by our algorithm are what we refer to as {\it maximal} RAF sets (maxRAFs). However, a maxRAF could possibly consist of several smaller (independent or overlapping) subsets which themselves are RAF sets (subRAFs). If such a subRAF cannot be reduced any further without losing the RAF property, we refer to it as an {\it irreducible} RAF (irrRAF). In \cite{Hordijk:04} we presented an extension of the RAF algorithm to find one (arbitrary) irrRAF within a given maxRAF. These notions of subRAFs are relevant for what follows below.

\section{The structure of autocatalytic sets}

In the original argument for the existence of autocatalytic sets in the binary polymer model, it was assumed that when the probability of catalysis $p$ is slowly increased (for a given $n$), at some point an autocatalytic set will occur as a ``giant component'', i.e., containing all (or most of) the reactions in the reaction set $\R$ \cite{Kauffman:71,Kauffman:86,Kauffman:93}. This is similar, it was claimed, to the sharp phase transition observed in percolation networks or random graphs. Furthermore, the proof that only a linear growth rate in the level of catalysis $f$ is sufficient to get RAF sets with arbitrary high probability assumes that RAF sets contain the entire molecule set $X$ \cite{Mossel:05}. However, as was already shown in \cite{Hordijk:04,Hordijk:11}, in practice these assumptions are stronger than necessary. Here, we investigate this issue of the size and structure of RAF sets in more detail.

\subsection{Empirical results}

\begin{figure}[htb]
\centering
\includegraphics[scale=.4]{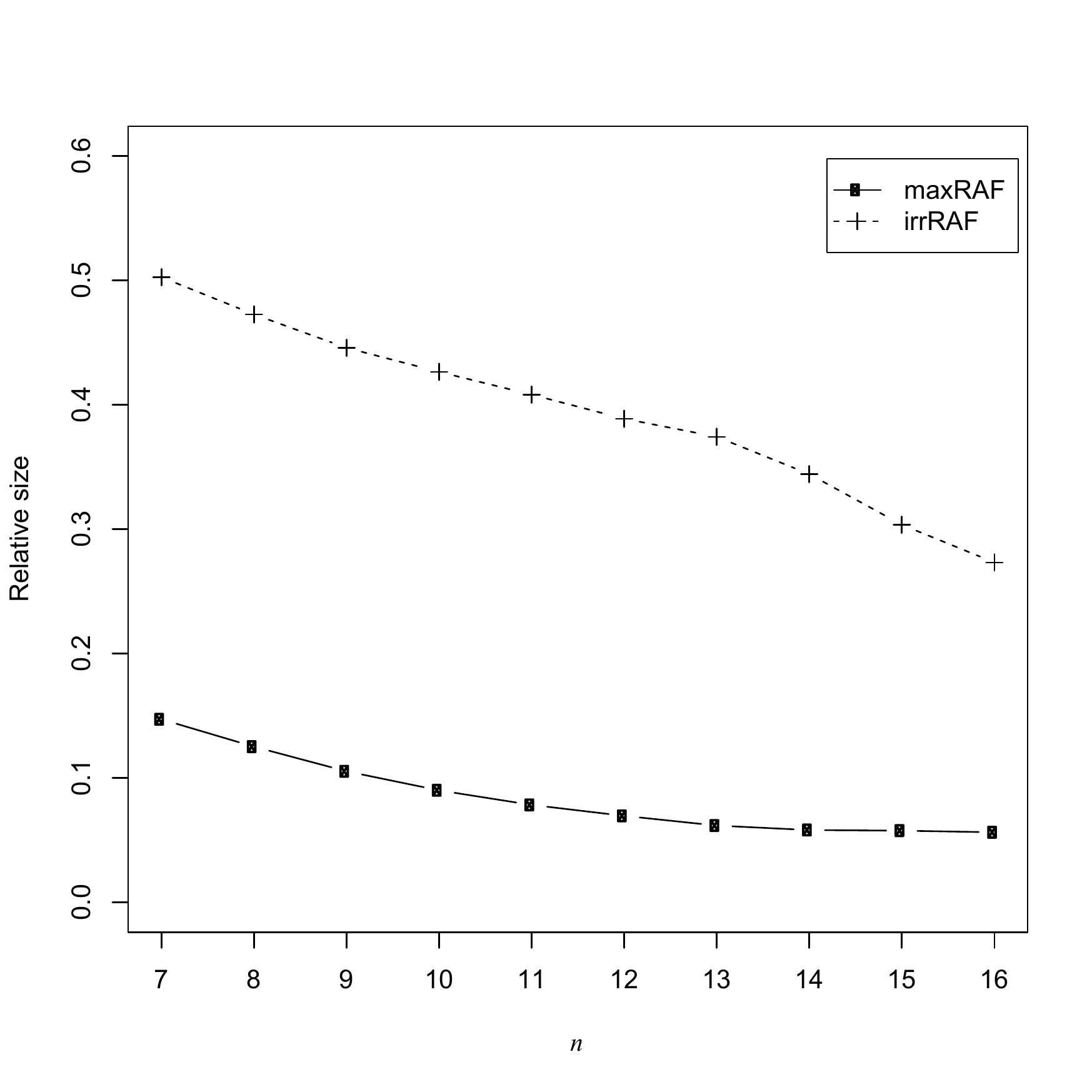}
\caption{The (average) relative size of maxRAFs and irrRAFs for increasing $n$ and $P_n \approx 0.5$.}
\label{fig:RAFsize_rel}
\end{figure}

First, we look at the {\it relative} size of RAF sets. For maxRAFs, we calculate the relative size as the size (in number of reactions) of a maxRAF divided by the size of the full reaction set $\R$ it was found in. For irrRAFs, we calculate the relative size as the size of an irrRAF divided by the size of the maxRAF it is part of. For various values of $n$ in the binary polymer model, we measured these relative sizes of maxRAFs and irrRAFs at a level of catalysis for which the probability of finding (max)RAF sets (using our RAF algorithm) is around $P_n=0.5$. We then averaged this over a sample of instances of the model (given the practical computational limitations, as the size of the reaction set $\R$ increases exponentially with $n$). Figure \ref{fig:RAFsize_rel} shows the results.

This plot clearly shows that maxRAFs are not necessarily giant components. In fact, for smaller values of $n$ the maxRAFs are only about 15\% of the size of the full reaction set, decreasing to about 6\% for $n=16$. Furthermore, irrRAFs tend to be about half the size of a maxRAF for smaller values of $n$, quickly decreasing to close to one quarter for $n=16$ (with a continuing decreasing trend). This seems to imply that maxRAFs indeed consist of multiple (possibly overlapping) irrRAFs, plus perhaps a number of reactions that are part of the maxRAF but not necessarily of any irrRAF.

\subsection{Theoretical results}

\begin{figure}[htb]
\centering
\includegraphics[scale=.35]{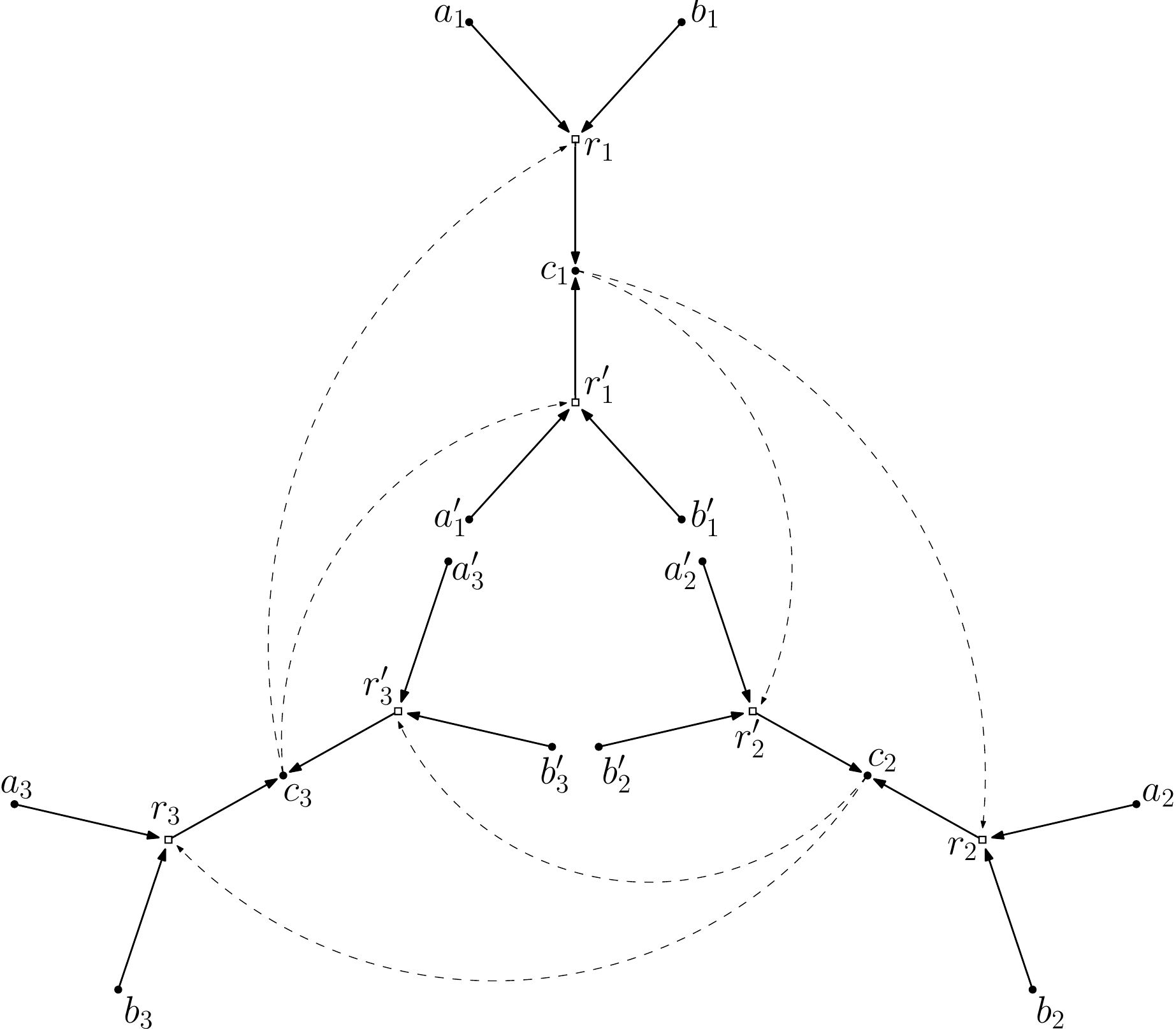}
\caption{An RAF set containing eight (overlapping) irreducible RAF sets, which illustrates the type of construction used to create $2^k$ irrRAFs (for $k=3$) in the proof of Theorem \ref{catlem}(1).}
\label{fig:exp_irrRAF}
\end{figure}

Next, we state several mathematical results (see Theorem \ref{catlem} below) concerning the structure of RAF sets, which confirm and formalize the implications that follow from the empirical results. The first result shows that an RAF set may indeed contain many irreducible RAF sets, in fact, possibly exponentially many (see Fig. \ref{fig:exp_irrRAF} for a simple example). One immediate consequence of this is that there is no hope of devising a polynomial-time algorithm that can guarantee to find {\it all} irrRAFs in an arbitrary RAF set (though an interesting, but still open question is whether irrRAFs can be merely counted in polynomial time, or can be listed by an algorithm which runs in polynomial time in the number of irrRAFs and the size of the RAF set). We will return to this issue with another, but more positive consequence (from an evolutionary point of view) with the example in Section \ref{sec:example}.  

For the second part  we first need to introduce one more definition. A {\it maximal proper subRAF}{\ } $\R''$ of an RAF $\R'$ is a proper subRAF $\R'' \subset \R'$ (i.e., $\R''$ is an RAF and is contained in but not equal to $\R'$) such that there is no other proper subRAF $\R^* \subset \R'$ with $\R' \subset \R^* \subset \R''$. Part 2 of Theorem \ref{catlem} shows that the number of maximal proper subRAFs of an RAF can only grow linearly with the size of the RAF. Thus, although an RAF can have many {\it minimal} (i.e., irreducible) proper subRAFs, it cannot have too many {\it maximal} proper subRAFs.

The third part of Theorem \ref{catlem} includes three related points. The first point will be used (in Section \ref{subsec:vis}) to show how we can represent all the subRAFs of an RAF by constructing a (Hasse) diagram; this provides a convenient way to visualize how the subRAFs sit inside each other. Next, we show there is an efficient (polynomial-time) method to determine whether any given RAF set can be decomposed into two (overlapping or disjoint) subRAFs. Together these two results provide a way to describe the possible pathways through which any particular RAF set might be built up by allowing subRAFs to combine in pairs, or to co-opt other reactions (see also the example in Section \ref{sec:example}). As a third point, we show that it is possible to efficiently determine whether or not any particular reaction or, more generally, any non-empty set of reactions, is ``essential'' within any given RAF, in the sense that {\em any} subRAF of the original RAF must contain this given set of reactions. 

The results described above are stated formally in the following theorem, the proof of which is provided in the Appendix.
\begin{theorem}
\label{catlem}
\mbox{ }
\begin{enumerate}
\item There exist RAF sets $\R'$ for which the number of irreducible RAF subsets is exponential in the number of molecules and reactions in $\R'$. 

\item For any RAF set $\R'$,  the number of maximal proper subRAFs of $\R'$ can never exceed $|\R'|$. 

\item Given a catalytic reaction system, $Q= (X, \R, C)$, a food set $F \subset X$, and an RAF $\R' \subseteq \R$, there exist polynomial time (in $|Q|$) algorithms that solve the following problems:
  \begin{itemize}
  \item[(i)] generate a list of all the maximal proper subRAFs of $\R'$;
  \item[(ii)] determine whether or not $\R'$ is the union of two proper subRAFs, and if so find all such pairs of subRAFs;
  \item[(iii)] for any given non-empty subset of $\R'$, determine whether that subset is contained in {\it every} subRAF of $\R'$.
  \end{itemize}
\end{enumerate}
\end{theorem}

\subsection{An illustrative example}  \label{sec:example}

\begin{figure}[htb]
\centering
\includegraphics[scale=.6]{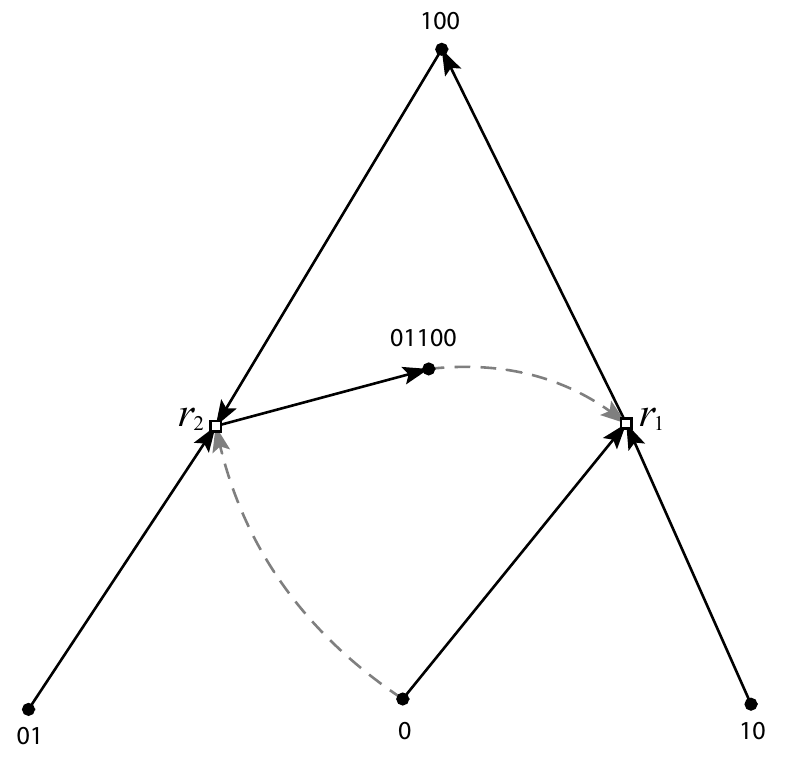}
\includegraphics[scale=.6]{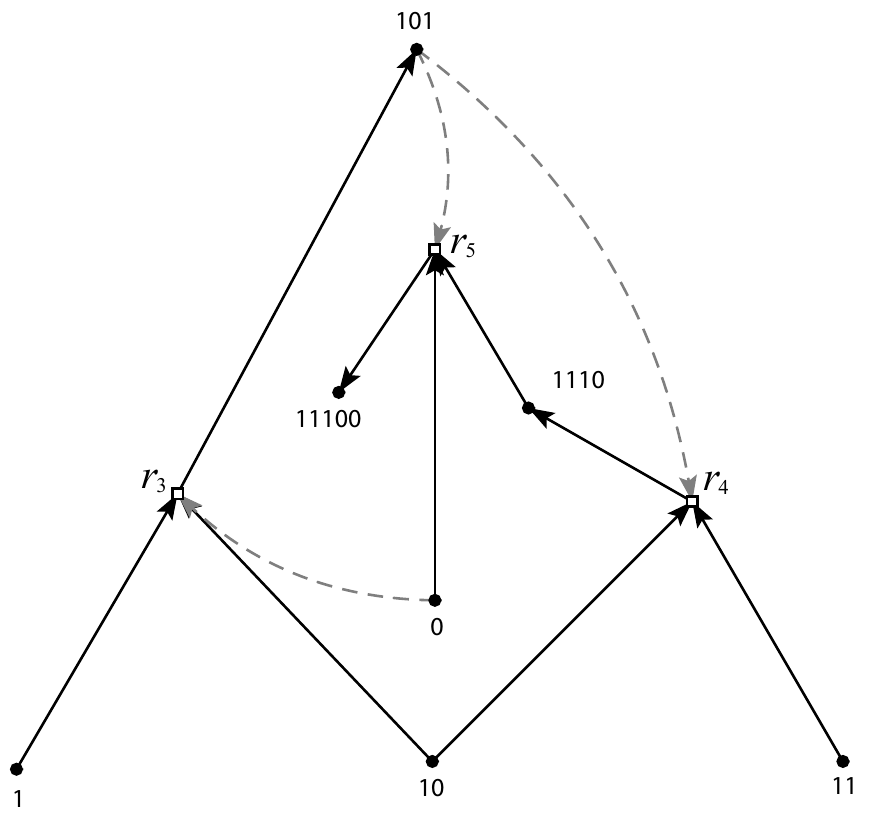}
\caption{The two independent subRAFs $\R_1$ (left) and $\R_2$ (right) of which the maxRAF shown in Figure \ref{fig:RAF_1} is composed.}
\label{fig:subRAFs}
\end{figure}

To illustrate this idea of the decomposition of an RAF set into smaller subsets, we provide a simple but illustrative example. Recall the maxRAF shown in Figure \ref{fig:RAF_1}, which consists of the following five reactions (the catalysts are shown above the reaction arrows):
\[ \begin{array}{llcl}
r_1: & 10 + 0   & \stackrel{01100}{\longrightarrow} & 100 \\
r_2: & 01 + 100 & \stackrel{0}{\longrightarrow}     & 01100 \\
r_3: & 10 + 1   & \stackrel{0}{\longrightarrow}     & 101 \\
r_4: & 11 + 10  & \stackrel{101}{\longrightarrow}   & 1110 \\
r_5: & 1110 + 0 & \stackrel{101}{\longrightarrow}   & 11100 \\
\end{array} \]
The food set is $F=\{0,1,00,01,10,11\}$, i.e., all molecules up to length $t=2$. This maxRAF can actually be decomposed into two independent subsets, both of which are RAF sets themselves. The first subRAF is $\R_1=\{r_1,r_2\}$ and the second subRAF is $\R_2=\{r_3,r_4,r_5\}$. These two subRAFs are shown in Figure \ref{fig:subRAFs}.

The subRAFs $\R_1$ and $\R_2$ are independent in the sense that they do not share any reactions. The only overlap is in the two food molecules $0$ and $10$ which are used in both subRAFs as reactants and a catalyst. Note, furthermore, that subRAF $\R_1$ is an irreducible RAF. None of the two reactions $r_1$ or $r_2$ can be removed without losing the RAF property. However, subRAF $\R_2$ is not an irrRAF, as first $r_5$ and then $r_4$ can be removed without losing the RAF property, leaving an irrRAF consisting of just one reaction ($r_3$, in which both reactants and the catalyst are food molecules). We discuss further details and implications of this example in section \ref{sec:evol} below.

\subsection{Classifying autocatalytic subsets}  \label{subsec:vis}

\begin{figure}[htb]
\centering
\includegraphics[scale=.7]{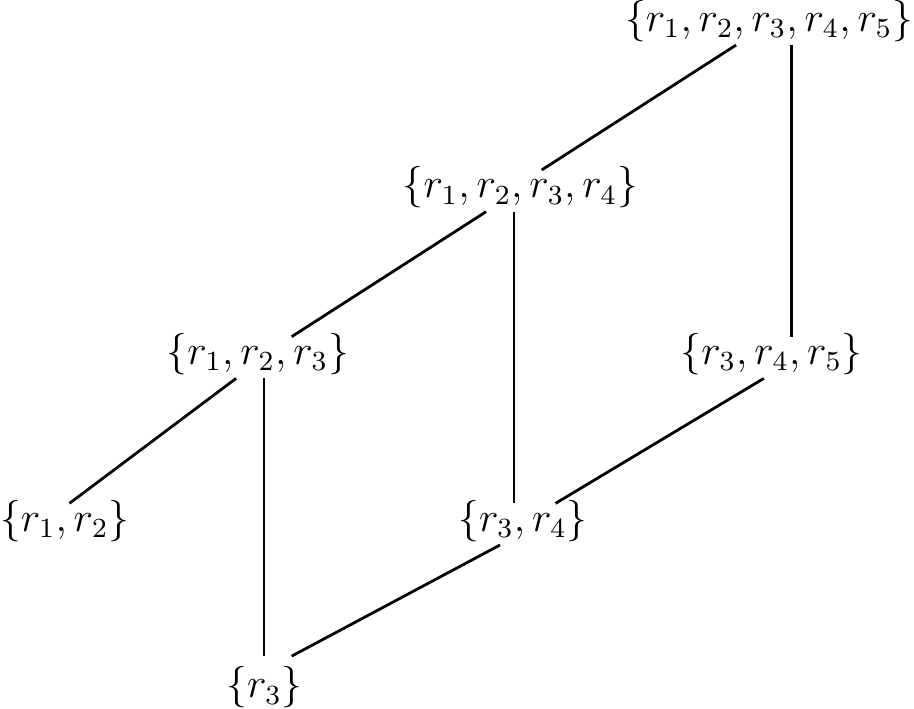}
\caption{The Hasse diagram of the poset of subRAFs of the maxRAF shown in Figure \ref{fig:RAF_1}.}
\label{fig:RAF_poset}
\end{figure}

The collection of subRAFs of any RAF $\R$ form a partially ordered set (i.e., a {\em poset}) under inclusion, and it is convenient to visualize this poset by its associated {\em Hasse diagram} \cite{Cameron:95}. This is a directed graph whose nodes are labeled by subRAFs of $\R$, with the top node (root) labeled $\R$ and the bottom nodes (leaves) labeled by the irrRAFs of $\R$. In this diagram we place an (upward oriented) edge from node $\R''$ to node $\R'$ precisely if $\R''$ is a maximal proper subRAF of $\R'$. For example, the Hasse diagram of the subRAFs of the 5-reaction maxRAF (from Figure \ref{fig:RAF_1}) is shown in Figure~\ref{fig:RAF_poset}. This Hasse diagram displays all the possible ways that the maxRAF can be built up from simpler subRAFs, starting from one or more irrRAFs.

In general, the poset of subRAFs of an RAF set can be very large. For example, as we have shown in Theorem~\ref{catlem}(1), an RAF set may have exponentially many irrRAFs, and each will appear as a separate node at the bottom of the Hasse diagram. Thus, there is no algorithm for constructing the Hasse diagram of the subRAFs that is guaranteed to run in polynomial time in $|Q|$ for all cases. However, Theorem~\ref{catlem}(3(i)) provides us with the next best thing -- namely, an algorithm that will be fast when the Hasse diagram is not too large. We formalize this as follows; a proof is provided in the Appendix.

\begin{corollary}
\label{catlem2}
Given a catalytic reaction system, $Q= (X, \R, C)$, a food set $F \subset X$, and an RAF $\R' \subseteq \R$, there is an algorithm for constructing the Hasse diagram of the poset $P$ of subRAFs of $\R'$ whose running time is polynomial in the size of $P$ and $\R'$.
\end{corollary}

\section{Evolvability, enablement, and emergence} \label{sec:evol}

The ``decomposability'' of RAF sets, as detailed in the previous section, actually gives rise to a possible mechanism for evolution to happen in autocatalytic sets. For example, initially the two subsets $\{r_1,r_2\}$ and $\{r_3\}$ in the above example could exist independently as irreducible autocatalytic sets. Over time, reaction $r_4$ and then reaction $r_5$ could be ``incorporated'' into the second subset, giving rise to the two independent subRAFs $\R_1$ and $\R_2$, which could then eventually merge into one larger RAF set consisting of all five reactions (as in Figure \ref{fig:RAF_1}). This process of combining, splitting, and recombining different subRAFs (presumed to be compartmentalized replicating entities) can give rise to inheritance, mutation, and competition, i.e., indeed \textit{evolvability}, as was convincingly shown only very recently \cite{Vasas:12}. It is in this context that a possibly exponential number of irrRAFs that can exist within a given maxRAF has an important (and positive, in terms of evolvability) consequence. Thus, our results above combined with those of \cite{Vasas:12} open up new and exciting possibilities, and warrant further investigations into the structure of (sub)RAFs under various conditions.

Furthermore, next to evolvability, the example from the previous section also illustrates how RAF (sub)sets can {\it enable} their own growth or even each other's coming into existence. Returning to the subRAF $\R_2$ in Figure \ref{fig:subRAFs} and the irrRAF $\{r_3\}$ it contains, suppose we start with only molecule types $0$, $1$, and $10$ present (all of which are food molecules). Then the irrRAF $\{r_3\}$ is a self-sustaining autocatalytic set by itself. If we now add molecule type $11$ (another food molecule) to the mix, then this minimal autocatalytic set immediately grows into a 2-reaction RAF, as reaction $r_4$ can now also proceed (given that it is catalyzed by molecule type $101$ which was already produced by reaction $r_3$). And, once the product of reaction $r_4$ (molecule type $1110$) is created, reaction $r_5$ is automatically added to the set as well, giving rise to the 3-reaction subRAF $\R_2$.

Finally, imagine that molecule type $11$ is not a food molecule, but some other molecule type that is actually produced by, for example, subRAF $\R_1$. Then the existence of $\R_1$ {\it enables} the growth and coming into existence of the full subRAF $\R_2$. Hence, (sub)RAF sets can be (possibly mutually) dependent on each other, and the existence of one can create the required conditions for another to come into existence (as was also already observed in \cite{Vasas:12}). Taking this one step further, one could imagine a collection of mutually dependent RAF sets forming a meta-RAF set: one set enabling (catalyzing) the existence of another, in mutually beneficial ways. In other words, self-sustaining, functionally closed structures can arise at a higher level (an autocatalytic set of autocatalytic sets), i.e., true {\it emergence}. And this, in turn, opens up the possibility of {\it open-ended} evolution.

\section{A generalized theory of autocatalytic sets}

We end this paper by discussing an as yet speculative, but potentially powerful and far-reaching idea. By definition, RAF sets are self-sustaining entities (supported by a food set) that exhibit catalytic closure. As we have argued here (largely inspired by \cite{Vasas:12}), they can give rise to evolvability, by virtue of their property of being decomposable into (possibly exponentially many) subsets. As such, the Hasse diagram of a (maximal) RAF set, as described above, can be interpreted as all the possible ``paths'' that evolution could potentially follow by combining, splitting, and recombining RAF subsets into many variants. Furthermore, as also argued here, RAF sets can enable their own growth, or even the coming into existence of other RAF sets, creating mutually dependent collections of RAF sets, and possibly emergent meta-RAF sets, thus giving rise to open-ended evolution.

So, if collections of molecules and chemical reactions between them, ``facilitated'' by catalysis, can form autocatalytic sets, perhaps even at several emergent levels, then this begs the question: ``Could we consider a complete cell as an (emergent) autocatalytic set?'' We argue that this may indeed be the case, particularly for autotrophs. And once autotrophs existed, they enabled the coming into existence of heterotrophs, i.e., collections of mutually dependent cells that feed on each other's waste or side products (or simply on each other). Perhaps it is not too far-fetched to think, for example, of the collection of bacterial species in your gut (several hundreds of them) as one big autocatalytic set.

Taking this a step further, why not consider any ecology of mutually dependent organisms as an emergent autocatalytic set, with one (group of) species enabling the evolution of (i.e., {\it creating niches for}) other, new, species. And what about the economy? If we view the process of transforming raw materials (reactants) into products as a ``production function'' (the equivalent of a chemical reaction), with objects like hammers, conveyor belts, and factory machines as ``facilitators'' (the equivalent of catalysts, which themselves are products of other production functions), perhaps we can also view the economy as an (emergent) autocatalytic set, exhibiting some sort of functional closure. And, as with an ecology, the existence of one or more autocatalytic subsets (economic agents such as companies or government organizations) enables the coming into existence of new ones.

We admit, once more, that all this is perhaps rather speculative, but at the same time we believe that these ideas are worth pursuing and developing further. In fact, the theory of autocatalytic sets, as we have only begun to formulate in the context of the origin of life, could perhaps be generalized into a theory of functional organization, and possibly also of emergence (collections of autocatalytic sets forming meta-autocatalytic sets). If so, this would open up multiple new areas of research, with many exciting prospects. As such, a ``generalized theory of autocatalytic sets'' might indeed fulfill such a promise.

\begin{acknowledgements}
This paper was finalized while WH and SK were visiting the Computational Systems Biology Research Group of the Tampere University of Technology, Finland. MS thanks the Royal Society of New Zealand for funding support. We also thank Vera Vasas for helpful and stimulating discussions.
\end{acknowledgements}

\bibliographystyle{spmpsci}
\bibliography{genRAF}

\section*{Appendix}

\textit{Proof of Theorem~\ref{catlem}:}

\bigskip

\textit{Part 1:} First, consider a directed graph $G$ that has $2k$ vertices $r_1, r_2,..., r_k$, and $r'_1, r'_2,..., r'_k$. For each $i=1,2,\ldots, k-1$, place a directed edge from $r_i$ to $r_{i+1}$ and also one from $r_i$ to $r'_{i+1}$. Next, for each $i=1,2,..., k-1$, place a directed edge from $r'_i$ to $r_{i+1}$ and also one from $r'_i$ to $r'_{i+1}$. Finally place directed edges from $r_k$ back to $r_1$ and to $r'_1$; similarly place directed edges from $r'_k$ back to $r_1$ and to $r'_1$.

Notice that the number of minimal directed cycles in this digraph is $2^k$, since we have complete freedom to select $r_i$ or $r'_i$ at each step in the cycle, and we must select one of them (to get a cycle) but not more than one (to get a minimal cycle).

We now use this graph to construct a RAF set that has exponentially many irrRAFs as follows. Associate with $r_i$ the reaction $a_i+b_i \rightarrow c_i$ and with $r'_i$ the reaction $a'_i + b'_i \rightarrow c_i$, where:
\begin{itemize}
\item[(i)] the $a_i, b_i, a'_i, b'_i$ and $c_i$  are all distinct from each other (and across different choices of $i$ there is no repetition), and
\item[(ii)] the $a_i, b_i, a'_i, b'_i$ are all in the food set $F$ (for all $i$).
\end{itemize}
For the catalysis set $C$, we let $c_i$ catalyze $r_{i+1}$ and $r'_{i+1}$ (for $i=1,2, \dots, k-1$). In addition, let $c_k$ catalyze $r_1$ and $r'_1$. Figure \ref{fig:exp_irrRAF} illustrates this RAF set for the case $k=3$.

The irrRAFs in this resulting RAF set are now in one-to-one correspondence with the minimal directed cycles of the graph $G$ described above, and there are $2^k$ such minimal cycles, but only $2k$ reactions and $5k$ molecules. So, the number of irrRAFs is exponential in the size of the RAF set. Notice that this construction can be carried out within the binary polymer model.

\bigskip

\textit{Part 2:} For an arbitrary subset $\R'' \subseteq \R$, let $s(\R'')$ denote the (possibly empty) subset of $\R$ obtained by applying the RAF algorithm to $\R''$ and $F$, and let $\R''_{\neq \emptyset}$ be the set of reactions $r$ in $\R''$ for which $s(\R''-\{r\}) \neq \emptyset$. We first establish the following result:

\vspace{0.3cm}
\noindent{\em Claim 1:} If $\R'$ is any RAF, then $\R''$ is a maximal proper subRAF of $\R'$ if and only if
\begin{itemize}
\item[(a)] $\R'' = s(\R'-\{r\})$ for some reaction $r \in \R'_{\neq \emptyset}$, and
\item[(b)] $\R''$ is not strictly contained within any other set of type (a). 
\end{itemize}
 
\noindent To verify this claim, suppose that $A$ is a maximal proper subRAF of $\R'$. Then there is at least one reaction $r \in \R'-A$. Notice that, since $A \subseteq \R'-\{r\}$, $s(A)=A$ is a non-empty subset of $s(\R' -\{r\})$; moreover $s(\R'-\{r\})$ is a strict subRAF of $\R'$ since $s(\R'-\{r\})$ does not include $r$ while $\R'$ does. Thus, since $A$ is a maximal proper subRAF of $\R$ we have
$$A= s(A) = s(\R'-\{r\}),$$
and so (a) holds. Property (b) now follows by the maximality assumption. 

Conversely, suppose that (a) and (b) hold for $\R''$. Then $\R''=s(\R'-\{r\})$ is nonempty and so $s(\R'-\{r\})$ is a proper subRAF of $\R'$, and if it were not a maximal proper subRAF of $\R'$ then, from the first part of the proof $s(\R'-\{r\})$ would need to be strictly contained within $s(\R' - \{r'\})$ for some reaction $r' \in \R'_{\neq \emptyset}$, and this is impossible since we are assuming that (b) holds.

From Claim 1, the number of maximal proper subRAFs is at most the number of sets of the form $s(\R'-\{r\})$ for $r \in \R'$, and there are at most $|\R'|$ such sets across the possible choices of $r$ from $\R'$. 
 
\bigskip

{\em Part 3:} Part (i) follows directly from Claim 1, since the collection of RAF sets $\{s(\R'-\{r\}): r \in \R'_{\neq \emptyset}\}$ can be computed in polynomial time, and property (b) in Claim 1 can then also be checked in polynomial time.

Part (ii) also follows from Claim 1, since this shows that $\R'$ is union of two proper subRAF if and only if 
\begin{equation}
\label{crucial}
\R' = s(\R'-\{r_1\}) \cup s(\R'-\{r_2\})
\end{equation}
for some pair of distinct elements $r_1, r_2$ of $\R'_{\neq \emptyset}$.

From this, it is clear how to obtain a polynomial time algorithm: first construct the set $\R'_{\neq \emptyset}$, and, provided this set is non-empty, search for all pairs $r_1, r_2 \in \R'_{\neq \emptyset}$ for which Eqn. (\ref{crucial}) holds; for each such pair we can set $\R_i:=s(\R'-\{r_i\})$, for $i=1,2$ so that $\R' = \R_1 \cup \R_2$. If no such pair $r_1, r_2$ exists (or if $\R'_{\neq \emptyset}$ is empty), then report that $\R'$ cannot be decomposed further. This completes the proof of the part (ii).

For part (iii), it suffices to verify the following:

\vspace{0.3cm}
\noindent{\em Claim 2:} If $\R'$ is any RAF set and $\R_0$ is any non-empty subset of $\R'$ then $\R_0$ is contained within every subRAF of $\R'$ if and only if $s(\R'-\{r\}) = \emptyset$ for all $r \in \R_0$. \\*

\noindent To verify this claim, first suppose there exists $r \in \R_0$ with $s(\R'-\{r\}) \neq \emptyset$. Then $s(\R'-\{r\})$ is a subRAF of $\R'$ and yet RAF $s(\R'-\{r\})$ does not contain $\R_0$, since 
$s(\R'-\{r\})$ is a subset of $\R'-\{r\}$ and so does not does contain $r \in \R_0$. Conversely, suppose there exists a subRAF $\R''$ of $\R'$ which does not contain $\R_0$. Select any reaction $r \in \R_0-\R''$. Then $\R'' \subseteq s(\R'-\{r\})$ and so $s(\R'-\{r\}) \neq \emptyset$. This establishes Claim 2, as required, and completes the proof.

\bigskip

\textit{Proof of Corollary~\ref{catlem2}:}
The algorithm constructs the Hasse diagram from the top down, starting from the single node $\R'$. We apply Part 3(i) of Theorem~\ref{catlem}  to list all the maximal proper subRAFs of $\R'$, and then place edges from each of these to $\R'$ (if $\R'$ has no maximal proper subRAFs then $\R'$ is irreducible and we leave the node as it is). Now we repeat this step recursively on these subRAFs, introducing edges as before, and also identifying any two (or more) nodes labeled by the same subRAF. We continue in this way until the network can be extended no further, in which case all the nodes with no children comprise the set of irrRAFs of $\R'$.

The resulting network $N$ that we have constructed contains all the
nodes of the Hasse diagram of the poset (i.e. it contains all the subRAFs of $\R'$); moreover, the edge set is a subset of the edges in the Hasse diagram.
This last claim needs a short proof: if we have constructed an edge in $N$ from
$\R_1$ to $\R_2$, where $\R_1 \subset \R_2$ we need to show that there is no other path in $N$ from $\R_1$ to $\R_2$  via a sequence of increasing subRAFs (which would make the edge $(\R_1, \R_2)$ redundant). Suppose there were such a second path, and let $(\R_3, \R_2)$ be the last edge on this path. Then, referring to Claim 1 (in the proof of Part 2 of Theorem~\ref{catlem}), $\R_1 = s(\R_2-\{r\})$ would be strictly contained in $\R_3 = s(\R_2-\{r'\})$ for some reactions $r, r'$ and this is forbidden in allowing $\R_1$ to be selected as a maximal proper subRAF of $\R_2$. 

Thus, each edge in $N$ will be present as an edge in the Hasse diagram. Moreover, all edges in the Hasse diagram are present in $N$, for suppose that in the Hasse diagram there is an edge from 
$\R_1$ to $\R_2$, where $\R_1 \subset \R_2$.   Then $\R_1$ must be a maximal subRAF of $\R_2$ and so, by construction, the algorithm inserts an edge from $\R_1$ to $\R_2$ during the step at which the subRAF $\R_2$ and its maximal subRAFs are considered.  

In summary, we have verified that the algorithm described constructs exactly the Hasse diagram of subRAFs of $\R'$.

\end{document}